\documentclass{emulateapj}
\usepackage{amsmath, natbib, mathtools}
\begin{document}

\title{On the Rarity of X-Ray Binaries with naked Helium Donors}

\author{T. Linden\altaffilmark{1,2}, F. Valsecchi~\altaffilmark{3}, V. Kalogera\altaffilmark{3} }
\affil{$^1$ Department of Physics, University of California, Santa Cruz, 1156 High Street, Santa Cruz, CA, 95064}
\affil{$^2$Center for Particle Astrophysics, Fermi National Accelerator Laboratory, Batavia, IL 60510}
\affil{$^3$Center for Interdisciplinary Exploration and Research in Astrophysics (CIERA) \& Department of Physics and Astronomy, Northwestern University, 2145 Sheridan Road, Evanston, IL 60208}
\slugcomment{Accepted by ApJ}
\shortauthors{}

\begin{abstract}
The paucity of known High-Mass X-Ray Binaries (HMXB) with naked He donor stars (hereafter He star) in the Galaxy has been noted over the years as a surprising fact, given the significant number of Galactic HMXBs containing H-rich donors, which are expected to be their progenitors. This contrast has further sharpened in light of recent observations uncovering a preponderance of HMXBs hosting loosely bound Be donors orbiting neutron stars (NS), which would be expected to naturally evolve into He-HMXBs through dynamical mass transfer onto the NS and a common-envelope (CE) phase. Hence, reconciling the large population of Be-HMXBs with the observation of only one He-HMXB can help constrain the dynamics of CE physics. Here, we use detailed stellar structure and evolution models and show that binary mergers of HMXBs during CE events must be common in order to resolve the tension between these observed populations. We find that, quantitatively, this scenario remains consistent with the typically adopted energy parameterization of CE evolution, yielding expected populations which are not at odds with current observations. However, future observations which better constrain the underlying population of loosely bound O/B-NS binaries are likely to place significant constraints on the efficiency of CE ejection.
\end{abstract}

\slugcomment{FERMILAB-PUB-11-518-A-T}

\section{Introduction}
As a star in a binary system evolves off of the main sequence (MS) and radially expands, it can overflow its Roche lobe and begin transferring mass onto its companion. If this transfer proceeds on a timescale shorter than the timescale in which the accretor can achieve thermal equilibrium, a common envelope (CE) develops and the stars begin to orbit through a combined atmosphere. Energy is then transferred from the binary orbit to the CE through frictional forces and torques which unbind the CE gases~\citep{1993PASP..105.1373I, 2000ARA&A..38..113T}. This process has long been discussed as the dominant mechanism for forming tight binaries from widely separated systems hosting one or two massive stars; it is thought to produce many closely interacting binaries such as Cataclysmic Variables \citep{1976IAUS...73...75P} and X-Ray Binaries \citep{1986MNRAS.220P..13E, 1987ApJ...316L..25B}.

Unfortunately the hydrodynamics and long-term evolution during the CE phase are not well understood, making it difficult to accurately determine the efficiency of energy transfer from the binary orbit into the ejection of the CE. While advancements in computational hydrodynamics have elucidated several trends near the onset of the CE phase~\citep{1996ApJ...471..366R, 2000ApJ...533..984S, 2006astro.ph.11043T, 2008ApJ...672L..41R}, a simple parametrization must be employed in situations where a population of binaries is to be considered. To this end, \citet{1984ApJ...277..355W} incorporated the myriad uncertainties of CE evolution into a single parameter $\alpha_{CE}$, which governs the efficiency of transferring gravitational energy into the complete removal of the CE. Within this framework, a relation between the initial and final orbital separations can be written as:

\begin{equation}
\label{eq:CE}
\alpha_{CE}~\left[\frac{GM_{c}M_{a}}{a_{f}} - \frac{G(M_{e}+M_{c})M_{a}}{a_{i}}\right]~=|E_{bind}|
\end{equation}

where G is the gravitational constant, a$_i$ and a$_f$ are the initial and final orbital separations, M$_{c}$, M$_{e}$ and M$_{a}$ are the masses of the donor core, donor envelope, and accretor respectively, and E$_{bind}$ is the energy necessary to unbind the CE. The value of E$_{bind}$ includes not only the (negative) gravitational binding energy, but also terms relating to the (positive) thermal energy of the plasma gas, the ionization of H and He, and the disassociation of H$_2$~\citep{1994MNRAS.270..121H, 1995MNRAS.272..800H}. 

Another uncertainty entering the calculation of E$_{bind}$ is the definition of the core-envelope boundaries~\citep{2000A&A...360.1043D, 2010arXiv1009.5400L}. As a rough definition of the stellar core, we assume the boundary to occur at the radius where the mass fraction of H drops below a critical value $X_{\rm{min}}$. Here we set $X_{\rm{min}}$=0.1 and investigate the effect of changing this parameter in what follows.

\begin{deluxetable*}{cccccc}
\tabletypesize{\scriptsize}
\tablewidth{0pt}
\tablehead{
\colhead{Name} & \colhead{Spectroscopy} & \colhead{$M_{\rm{don}} (M_\odot$)}& \colhead{$P_{\rm{orb}}$ (d)} & \colhead{$e$}
}
\startdata
$\gamma$-Cas & B0.5 IVe & 13.8 & 203.59 & 0.26 \\
0115+634 & B0.2 Ve & 14.9 & 24.3 & 0.34 \\
0236+610 & B0.5 Ve & 13.8 & 26.5 & 0.55 \\
0331+530 & O8-9 Ve & 23.0 & 34.3 & 0.3 \\
0352+309 & O9.5 IIIe-B0 Ve & 16.7 & 250 & 0.11 \\
0535+262 & B0 IIIe & 15.6 & 111 & 0.47 \\
0834-430 & B0-2 III-Ve & 12.0 & 105.8 & 0.12 \\
J1008-57 & O9e-B1e & 15.6 & 247.5 & 0.66 \\
1417-624 & B1 Ve & 12.0 & 42.1 & 0.446 \\
J1946+274 & B0-1 IV-Ve & 13.8 & 169.2 & 0.33 \\
J1948+32 & B0 Ve & 15.6 & 40.4 & 0.03 \\
2030+375 & B0e & 15.6 & 46.03 & 0.41 \\
J2103.5+4545 & B0 Ve & 15.6 & 12.67 & 0.40 
\enddata
\tablenotetext{}{\label{tab:behmxbs} Table 1: List and parameters of the 13 galactic Be-HMXBs with complete binary orbital information. Data obtained from~\citet{2005A&AT...24..151R} and \citet{2009ApJ...707..870B}, and numerous references therein. The donor mass, orbital period and eccentricity are denoted by $M_{\rm{don}}$, $P_{\rm{orb}}$, and $e$, respectively. Each donor mass is derived from the spectral classification using Table VIII of \citet{1981A&AS...46..193H}. }
\end{deluxetable*}

One long-standing expectation from models of binary evolution concerns the resilient population of high-mass X-ray binaries (HMXB) consisting of neutron stars (NS) accreting matter from the wind of naked He donors (He-HMXB). Such binaries are expected to form when the observed population of O/B - HMXBs evolves through a CE phase of the supergiant stellar component. The lack of observed He-HMXBs is particularly puzzling in light of recent observations uncovering an unexpectedly large population of HMXBs with Be-type stars (B-type stars which show emission-line (Be) spectra). Since each Be-HMXB contains a NS accretor along with a massive (O8-B2) donor in a wide orbital period (P$_{orb}~>~$30~days), these systems are the expected progenitors of a bright HMXB population with He donors. Specifically, as both the orbital separation of observed Be-HMXBs is smaller than the star's supergiant radius and the mass ratio between the Be donor and NS is large, Be-HMXBs inevitably evolve into a CE. If the binary survives the CE event, the resultant system would contain a He star coupled with a NS in a relatively tight orbit. Since He stars can experience significant mass loss due to stellar winds ~\citep{1988A&AS...72..259D}, we would naively expect to observe such binaries as bright He-HMXBs. At least 81 galactic Be-HMXBs are currently observed, with 69 reported by both ~\cite{2005A&AT...24..151R}, \citet{2009ApJ...707..870B}, and numerous reference therein, as well as 12 additional systems recently discovered by INTEGRAL~\citep{2008MNRAS.386.2253K, 2009A&A...494..417R, 2010A&A...517A..14R, 2009ATel.2008....1C}. Of these systems, 13 have known orbital period, eccentricity, and spectral data, and are summarized in Table~\ref{tab:behmxbs}. Despite the large number of known Be-HMXBs, several decades of X-ray observations have uncovered only one He-HMXB, Cyg X-3~\citep{1992Natur.355..703V}. 

It is important to note that, throughout this paper, we specifically avoid the term Wolf-Rayet (WR) for theoretically modeled He stars, as the exact definition of a WR star varies throughout the literature. While observational studies focus on spectroscopic WRs (e.g. \citealt{2000A&A...360..227N}), many population synthesis investigations treat \emph{all} He stars as WRs, applying a WR-like wind mass-loss (e.g.~\citealt{2000MNRAS.315..543H} and \citealt{2008ApJS..174..223B}, who use the wind prescription from \citealt{1995A&A...299..151H, 1998A&A...335.1003H}). On the other hand, other theoretical studies have set a minimum mass for a WR in a HMXB of 5-8~M$_\odot$,  consistent with the lowest WR mass inferred spectroscopically \citep{1997A&A...318..812D, 1998NewA...3..443V, 2005A&A...443..231L}. The mismatch between the various definitions is further complicated by the fact that a low-mass He star ($<$~1.5M$_\odot$) undergoing Roche-Lobe overflow onto a compact object may replicate the spectroscopic characteristics of the higher mass observed WR population \citep{2005A&A...443..231L}. For this reason, in what follows we only call a star WR if it is a He star which shows spectroscopic characteristics typical of observed WR stars.

An analysis by \citet{1976IAUS...73...35V} estimates a population of He-HMXBs that exceeds the population of HMXBs with H-rich supergiant donors by a factor of 15, and argues that the lack of observed He-HMXBs provides an observational argument in favor of mergers during the dynamically unstable mass transfer phase. More recent population synthesis calculations by \citet{2005A&A...443..231L} find that the galaxy is expected to contain $\sim$1 HMXB powered by the stellar wind of a He donor star larger than 7M$_\odot$, as well as $\sim$1 HMXB powered by the Roche lobe overflow of a He star less massive than 1.5M$_\odot$. They conclude that it is equally probable that Cyg X-3 is either type of system. \citet{1998A&ARv...9...63V} also predict a significant population of binaries hosting a NS and a He donor star, but note that the propeller effect may prevent significant accretion onto the compact object. \citet{2000A&A...358..462V} note that the lack of HMXBs with He donors may be consistent with the lack of observed He stars in the solar neighborhood. Finally, \citet{2002MNRAS.331.1027D} examined the production of double-NS binaries from a population of low-mass He-HMXBs, finding that double-NS systems are only produced if the He star is undergoing He-shell burning during the period of RLO, while systems with He-core burning during RLO instead become white dwarf-NS binaries. However, they do not discuss the X-Ray characteristics of their He-HMXB population.

Additionally, a significant effort has focused also on the population of observed runaway WRs, which may be expected to host compact object partners. \citet{1982IAUS...99..577M,1982A&A...114..135M} present spectroscopic observations of such systems, but did not detect any X-ray bright source. Population synthesis studies targeting He stars more massive than 7~M$_\odot$ found that only a small portion (1-2\%) of the observed runaway binaries hosting an O-star and a He star may eventually evolve into He-HMXBs. Subsequent work by \citet{1998NewA...3..443V} found that at most 3\% of the galactic population of massive He stars should be found in binaries containing a compact object. Both studies determined the paucity of observed He-HMXBs to be primarily due to the effect of disruptions due to the natal kick imparted to the compact objects at formation, as well as the probability of mergers during CE events. However, these results cannot be directly applied to our study of He-HXMBs stemming from the observed Be-HMXB population, as we know that the system must survive as an intact binary through the NS natal kick. In any case, such population of runaway He stars is unlikely the evolutionary outcome of a population like the observed Be-HMXBs as the high spatial velocities of these systems are likely associated with the formation of the compact object and the observed O/B-HMXBs show significantly smaller spatial velocities \citep{1998A&A...330..201C, 2000A&A...364..563V}.

In the present analysis, we use detailed stellar structure and evolution models to investigate whether the discrepancy between Be-HMXB and He-HMXB observations can be used to place constraints on the dynamics of the CE event, and specifically on $\alpha_{CE}$. In \S~\ref{sec:modeling}, we describe the modeling codes developed to calculate the orbital parameters of XRBs immediately before and after the CE. In \S~\ref{sec:results}, we use the currently observed sample of Be-HMXBs with measured orbital periods and eccentricities, and find that detailed calculations of massive-star binding energies and the typically adopted energy parametrization of CE evolution are consistent with the observed Galactic sample of one He-HMXB. In \S~\ref{sec:grid}, we investigate whether further limits can be placed on $\alpha_{CE}$ by simulating a grid of widely separated O/B-NS binaries and determining the survival probability of He-HMXB systems as a function of $\alpha_{CE}$. In this way, we address the question of how a significant population of undiscovered O/B-NS binaries with wide orbits could further constrain energy deposition during the CE phase. We conclude in Section~\ref{sec:conclusions} with a discussion of how future observations could be used to place further constraints on $\alpha_{CE}$.

\section{Simulation and Modeling Codes}
\label{sec:modeling}

We have developed a detailed orbital evolution code suitable for investigating XRBs. This code tracks the evolution in time of the orbital separation and eccentricity of the binary, and the spin of the stellar component, accounting for the competing effects of stellar wind mass loss, wind accretion, tides exerted from the compact object onto the star, and angular momentum loss via gravitational radiation. To account for changes in the stellar properties due to natural stellar evolution, the orbital evolution code is coupled to a detailed stellar structure and evolution code. The stellar evolution models are calculated with an up-to-date version of the stellar evolution code STARS (\citealt{Eggleton1971, Eggleton1973, Eggleton2002, Polsetal1995}), while the NS is considered as a 1.4~M$_\odot$ point mass.  With this code, we follow the orbital evolution of Be-XRBs from the formation of the NS to the onset of Roche-lobe overflow (RLO), which corresponds to the beginning of the CE phase. At this stage we extract from the stellar model the donor's parameters (e.g. core and envelope masses, envelope binding energy, and spin) as well as the computed orbital separation and eccentricity.

Tidal evolution is calculated in the standard weak-friction approximation (\citealt{Zahn1977, Zahn1989}), following the formalism of \cite{Hut1981}. Specifically, we integrate numerically the set of differential equations as presented in \S~3.3 of \cite{Belczynski2008}, with the only modification being in the second-order tidal coefficient $E_2$. For this coefficient, we adopt a stellar model from \cite{2004A&A...424..919C} with a mass of $\sim16\,$M$_\odot$ at solar metallicity and derive E$_2$ as:

\begin{equation}
\log{E_2} = -5.71 -2.51\cdot t\rm_{MS}^{17284.8}-0.69\cdot t\rm_{MS}^{74.24} -2.30\cdot t\rm_{MS}^{2.10}
\end{equation}

where $t\rm_{MS}$ is time in units of the main sequence lifetime. The fitting formula has only a very weak dependence on the initial mass in the observed range for the Be donor mass. The evolution of the orbital separation and eccentricity driven by stellar wind are calculated following \cite{2002MNRAS.329..897H}. If the primary loses mass $\Delta M_1$, the orbit loses $\Delta M_1 R_1^2\Omega_{orb}$ of angular momentum. If the secondary accretes some of this mass $\Delta M_2$, then $\Delta M_2R_1^2\Omega_{orb}$ is returned to the orbit, where $\Omega_{orb}$ is the orbital frequency and $R_1$ is the radius of the primary. The time evolution of the rotational frequency driven by stellar wind follows \cite{2000MNRAS.315..543H} and we assume that all the mass is lost uniformly at the surface of the star. The evolution of the orbital separation and eccentricity due to gravitational radiation is calculated following \cite{Junker1992}. The accretion efficiency is calculated according to \cite{BondiHoyle1944}  following \S~4.2 of \cite{Belczynski2008}. 

For each time step during the orbital evolution we compute the Roche-lobe radius of the star at periastron (\citealt{SepinskyEtAlNoFred2007}) and halt the calculation at the onset of the CE phase.  At this stage we extract the relevant binary and stellar parameters (see Eq.~\ref{eq:CE}) and calculate the resultant orbital parameters of the binary system from the energy prescription for CE evolution. If the radius of the donor star is found to exceed its Roche Lobe after the application of the CE prescriptions, we consider the system to end in a binary merger. 

Here we note that the mass-loss rate associated with the wind of a He star is uncertain and this, in turns, affects the X-Ray detectability of any surviving He-NS binary. Two very different models are quoted in the literature. First, following the models of \citet{2000MNRAS.315..543H} we assign a single power law to the mass loss spectrum for all He stars stemming from massive binaries, given by:

\begin{equation}
\dot M_{He}=1\times10^{-13} (L/L_\odot)^{1.5}
\label{windHurley}
\end{equation}

However, work by \citet{2000A&A...360..227N} and \citet{2001A&A...376..950N} have produced a broken power law where the mass loss declines precipitously for lower-mass He stars. The best fit given by \citet[][Eq. 6]{2002MNRAS.331.1027D} follows:

\begin{equation}
\dot M_{He}= 
\begin{Bmatrix*}
2.8\times10^{-13}(L/L_\odot)^{1.5} ~~~ log(L/L_\odot)> 4.5\\
4.0\times10^{-37}(L/L_\odot)^{6.8} ~~~ log(L/L_\odot)<4.5
 \end{Bmatrix*}
 \label{windDewi}
\end{equation}

These models yield very different X-ray luminosities for the case of the lower mass He-HMXBs which are expected to result from the observed Be-HMXBs. In this work, we use both Eqs.~(\ref{windHurley}) and (\ref{windDewi}) and 
assess the effect on the observability of the He-HMXB population. After applying these wind-mass loss rates, we follow \citet[Eqs. 39 \& 83]{2008ApJS..174..223B} to calculate the X-Ray luminosity of each individual system and to determine whether it enters the $Chandra$ band.

Another uncertainty in this binary modeling concerns the evolutionary stage at which the Be-HMXB is observed, which affects how long the orbital evolution calculation should persist before the Be donor begins RLO. In order to constrain the error introduced by this uncertainty, we test three scenarios where current Be-HMXBs are assumed to be observed at ZAMS, terminal-age main sequence (TAMS), and an intermediate case in the middle of the main sequence lifespan (MAMS). This effectively brackets the uncertainty stemming from the evolutionary state of the observed Be-HMXB population. We note that the long orbital periods of Be-HMXBs rule out the possibility that this MS evolution is interrupted by CE phases prior to the formation of the NS component, as CEs are known to produce binaries with substantially shorter orbital periods than observed in any Be-HMXB system~\citep{2011arXiv1109.6662N}.

In addition, we investigate the production of He-HMXBs from currently observed Be-HMXBs by expanding on this procedure through the creation of a grid containing systems with parameters similar to the observed Be-HMXB population. Specifically, we use the detailed stellar evolution code STARS to create a grid of stellar models with masses between 10-30M$_\odot$ (with a resolution of 1M$_\odot$) and a probability distribution following \citet{1955ApJ...121..161S}. We assume an initial orbital period in the range 10-200d with a resolution of 1 day, and a density distribution which is flat in the logarithm of orbital period, which we use as a tracer for the orbital separation~\citep{1983ARA&A..21..343A}. In an alternative model we also investigate systems with orbital periods extending out to 400d and  1000d, to determine the impact of this cutoff on our results. Finally, we employ a thermal eccentricity distribution following \citet{1975MNRAS.173..729H}.  From this grid, the probability of a given progenitor surviving to become a visible He-HMXBs can be computed for an arbitrarily large population of likely progenitor systems. The normalization of the survivable probability to stellar environments can be ascertained through normalization against the observed number of Be-HMXBs as described in Section~\ref{sec:grid}.

\section{Results for the Observed Be-HMXBs}
\label{sec:results}

\begin{figure}
\plotone{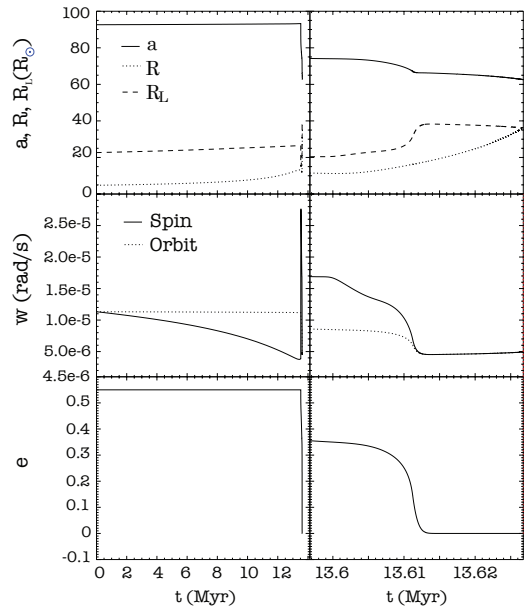}
                \caption{Evolution of the orbital parameters for a binary with the properties of 0236+610 (see Table~\ref{tab:behmxbs}). \textit{Top}: orbital separation ($a$), radius ($R$), and Roche-lobe radius ($R_L$); \textit{middle}: spin and orbital angular frequency ($w$); \textit{bottom}: orbital eccentricity ($e$). The left panel shows the overall evolution, while the right panel shows a zoom-in at RLO.  For the first $\simeq13\,$Myr, stellar wind mass-loss drives the evolution of the orbital separation, while the expansion of the star during its MS lifetime causes a decrease of the spin angular velocity. The evolution of the orbital eccentricity is driven by tides, which are too weak to affect it significantly. Towards the end of the MS the development of a convective envelope greatly increases the efficiency of tides. This mechanism controls the remaining orbital evolution leading to a decrease in the orbital separation and eccentricity, and driving the system into spin-orbit synchronism.}
\label{fig:DetailedOrbitalEv_0236_610}
\end{figure}

In Figure~\ref{fig:DetailedOrbitalEv_0236_610}, we illustrate the dynamics of our orbital evolution code by providing the detailed evolutionary history for a single simulation of 0236+610 (Table~\ref{tab:behmxbs}). Specifically, we plot the evolution in time of the orbital separation, eccentricity, stellar radius, Roche-lobe radius, spin and orbital frequencies until the onset of RLO, when our calculation is halted. In this simulation we assume that the Be companion is currently observed at ZAMS.  

From this detailed analysis it is evident that the evolution of the each parameter is dominated by only the last 0.1\% of its total lifetime, when the star develops a convective envelope which strongly enhance the tidal evolution. This suggests that simulations beginning at ZAMS, MAMS or TAMS will produce nearly identical binary parameters at the onset of the CE. 

The currently observed population of Be-HMXBs includes only 13 systems with sufficient observational constraints to allow for a detailed modeling of their orbital evolution forward in time (listed in Table~\ref{tab:behmxbs}). Therefore, it is important to first determine which factors, other than CE dynamics, could prevent the observation of He-HMXBs. Taking a small step aside, we test two likely factors, the relative lifetime of a bright He-HMXB phase compared to the Be-HMXB phase, and the expected X-Ray luminosity of the He-HMXB population. 

\subsection{Alternative Explanations for the lack of Observed He-HMXBs}
\label{subsec:alternatives}
\begin{figure}
                \plotone{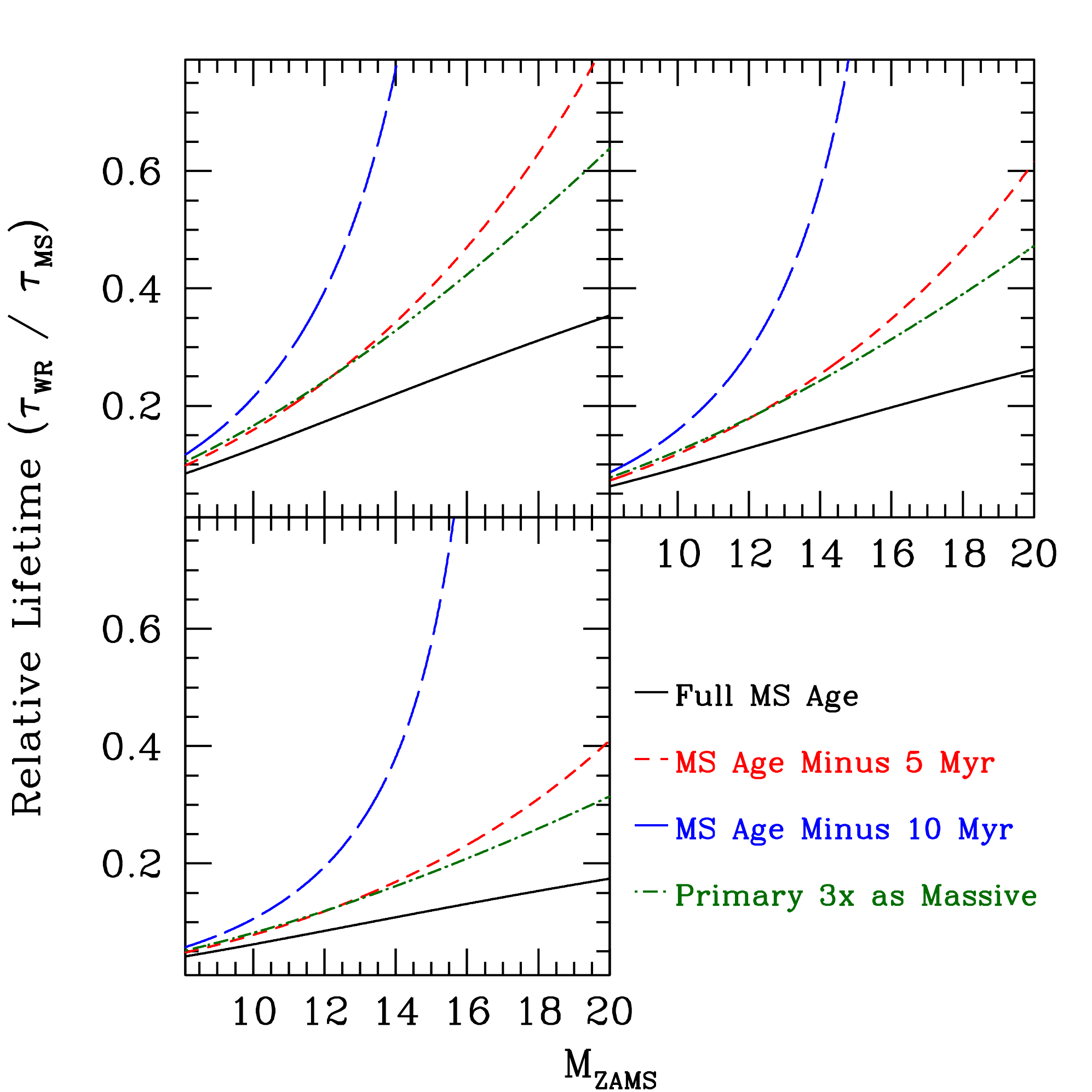}
                \caption{Relative lifetime of a 2.5~M$_\odot$ (top left), 3~M$_\odot$ (top right) and 4~M$_\odot$ (bottom left) He star compared to the MS lifetime of O and B stars with various initial masses. To translate these results into a comparison of He-HMXB lifetimes and Be-HMXB lifetimes, we make four different assumptions regarding the fraction of the MS lifetime during which the star is a Be-HMXB: it's entire lifetime (black solid), beginning at 5~Myr (red dashed), beginning at 10~Myr (blue long dashed), at the SN age of a primary star which is three times as massive as M$_{ZAMS}$ (green dot-dashed). The mass range of 2.5-4 M$_\odot$ for our population of He stars is set by the range of He core masses found in our models.}
\label{fig:relativetimescale}
\end{figure}

One possible explanation for the lack of observed He-HMXBs concerns the relative duration of He-HMXB and Be-HMXB phases. If the time the stellar component spends as a He star is only a small fraction of the Be-HMXB lifetime, we would be unlikely to observe a large population of these systems regardless of the CE dynamics. In order to investigate this effect, we follow the calculations of \citet{2000MNRAS.315..543H} where the lifetimes of the MS and He-MS phases ($\tau_{MS}$ and $\tau_{He}$ respectively) are given by:

\begin{equation}
\tau_{MS} = \frac{1594+ 2707M_0^4 + 146.6M_0^{5.5} + M_0^7}{0.04142M_0^2+0.3426M_0^7} Myr
\end{equation}

\begin{equation}
\tau_{He} = \frac{0.4129 + 18.81M_{He}^4 + 1.853M_{He}^6}{M_{He}^{6.5}} Myr
\end{equation}

where M$_0$ is the ZAMS mass and M$_{He}$ is the mass of the He main sequence star. We note that the direct comparison of these lifetimes sets an extremely conservative lower bound on the population of expected He-HMXBs, as it assumes that the Be-HMXB is X-Ray bright for the entire MS lifetime of the donor star. This assumption is unrealistic, as it includes the portion of the MS lifetime which occurs prior to the evolution of the primary star into a NS. In Figure~\ref{fig:relativetimescale}, we plot the fractional lifetime for a He star of 2.5M$_\odot$, 3M$_\odot$, and 4M$_\odot$ as a function of the ZAMS mass under four assumptions for the fraction of the total MS lifetime our observed systems spend as a Be-HMXB. First, we assume that the system exists as a Be-HMXB for it's entire MS lifetime. Secondly, we subtract 5~Myr from the MS lifetime to account for the formation of an NS from the most massive progenitors \citep[see e.g.][]{2006ApJ...636L..41M}. Third, we subtract 10~Myr from the NS lifetime to account for the average lifetime of NS progenitors~\citep{2003ApJ...591..288H}. 
Lastly, we follow the model of ~\citet{2005ApJS..161..118M}, who propose that the early evolution of Be-HMXBs is governed by stable mass transfer from the NS progenitor onto the Be progenitor. This scenario sets an upper limit on the mass ratio between the primary and secondary star at the onset of RLO based on the requirement that the accreting star achieves thermal equilibrium on a timescale smaller than the mass transfer timescale of the primary star. While the exact mass ratio may depend sensitively on the evolutionary state of each stellar component~\citep{1993PASP..105.1373I}, for MS companions \citet{1989PhDT.........7H} set a range of 2-4, while more recent work by \citet{2004ApJ...601.1058I} set a mass ratio of approximately 3. Motivated by these analyses, we lastly calculate the relative lifetime of the Be-HMXB phase as the MS lifetime of a Be star minus the MS lifetime of a progenitor star which is initially three times as massive. We note that all these scenarios are fairly conservative, due to the possibility that NSs in Be-HMXBs are formed via Electron-Capture supernovae~\citep{1984ApJ...277..791N}, which sets much stronger constraints on the lifetime of the Be-HMXB phase~\citep{2009ApJ...699.1573L}. We note that a range of 2.5-4~M$_\odot$ for the mass of the He core at the time of CE formation is strongly suggested by the results of our detailed stellar evolution models.

\begin{figure}
                \plotone{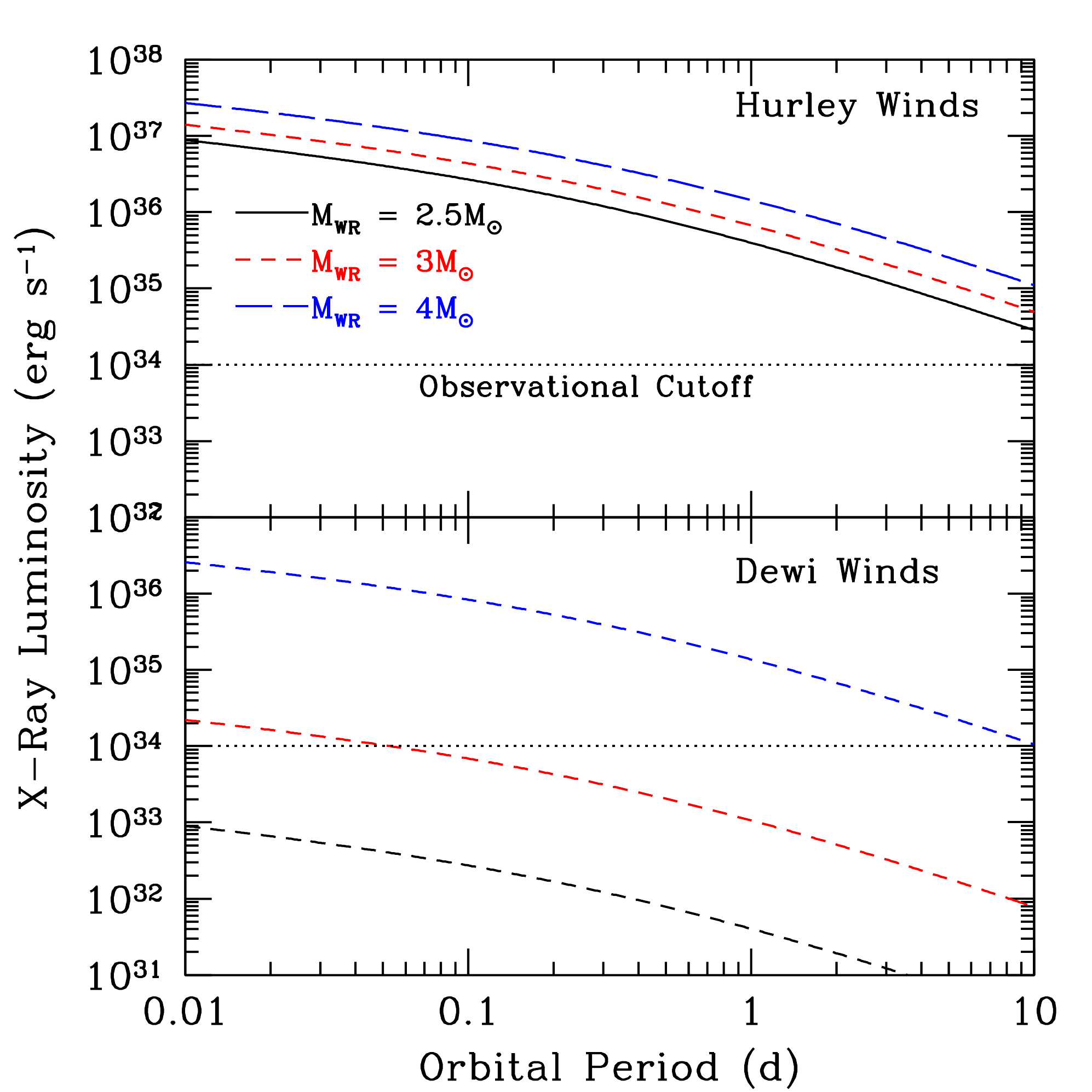}
                \caption{X-Ray Luminosity in the $Chandra$ band for a HE-NS HMXB as a function of the orbital separation, using the He wind prescriptions of \citet{2000MNRAS.315..543H} (top) and the He wind prescriptions of \citet{2002MNRAS.331.1027D} (bottom) for He star masses spanning the range of those produced via our stellar evolution code. In each case we use the X-Ray luminosity prescriptions of \citet{2008ApJS..174..223B}}
\label{fig:wrlum}
\end{figure}

Even considering the most massive He stars (with the shortest lifetime), as well as the most conservative calculation of the Be-HMXB lifetime, we expect a fractional lifetime ($\tau_{He}$/$\tau_{Be}$) of between 5-10\% for Be stars between 10-15 M$\odot$. Thus the population of 81 currently observed Be-HMXBs predicts a population of at least 6 He-HMXBs, which is at odds with the observation of only one such system at the 2$\sigma$ level. A significantly larger population exceeding 15 He-HMXBs is expected from more reasonable calculations of the luminous Be-HMXB lifetime and the He rich stellar mass. Thus, we may reject the hypothesis that the lack of observed He-HMXBs stems from their short lifespan.

A second explanation for the lack of observed He-HMXBs concerns their assumed X-Ray luminosities. If mass lost from the He star is not efficiently transferred onto the NS, the systems may simply fall below the luminosity threshold of present observations. In this work, we assume that any system with an X-Ray luminosity in the $Chandra$ band exceeding 1x10$^{34}$~erg~s$^{-1}$ would have been detectable in galactic surveys~(Andreas Zezas, Private Communication). We note that this assumption is conservative, and based primarily on the sensitivity of X-Ray survey missions such as ROSAT. A comparison with the X-Ray luminosities of the Be-HMXB population show several systems with detected luminosities below this level~\citep{2005A&AT...24..151R}.

We note that these calculations depend sensitively on both the mass of the He core determined by our evolutionary code, as well as the wind mass loss rate assigned to these systems. We find that our He stars span a mass range of approximately 2.5-4 M$_\odot$ with one outlier exceeding 6~M$_\odot$. In Figure~\ref{fig:wrlum} we plot the calculated X-ray luminosity as a function of the orbital period for He-HMXBs hosting a He stellar component with a mass of 2.5M$_\odot$ (black), 3M$_\odot$ (red), and 4M$_\odot$ (blue), and following the wind mass-loss models of both \citet{2000MNRAS.315..543H} and \citet{2002MNRAS.331.1027D} as given in Sec.~\ref{sec:modeling}. If stellar winds following the prescription of \citet{2000MNRAS.315..543H} are applied, these systems remain above the luminosity threshold out to orbital separations greatly exceeding those expected in post-CE binaries, which follow a logarithmic normal distribution spanning from 1.9h to 4.3 days, with a peak at 10.3h \citep{2011arXiv1109.6662N}. However, in the case of the weaker winds applied by \citet{2002MNRAS.331.1027D}, the final orbital separation of the systems may affect its detectability in the $Chandra$ band. Throughout what follows, we adopt the more conservative He mass-loss prescriptions of \citet{2002MNRAS.331.1027D}, and only count as ''detectable" those He-HMXBs with a luminosity exceeding 1$\times$10$^{34}$~erg~s$^{-1}$.

Another process that might prevent He-NS binaries to be detected as X-ray sources is the propeller effect. It has been previously noted \citep{1998A&ARv...9...63V} that a high rotational velocity for the NS accretor may prevent accretion of the donor wind material onto the NS. However, this is unlikely to affect the He-HMXB population modeled here. Since these systems are observed to be X-Ray bright during the Be-HMXB phase, when the winds are less intense and the orbits are significantly wider, it is unlikely that the propeller effect will prevent the much stronger wind accretion during the tighter, post-CE phase. While the NS may be spun up during the CE phase, this spin-up is thought to be accompanied by a decrease in the NS magnetic field. Such an effect is indeed observed for the pulsars contained in double-NS systems which have moved through CE evolution. These systems are expected to exist as the offspring of the He-HMXB population, and their reduced magnetic fields allow at least quasi-periodic accretion onto the NS despite the high angular momentum in these NS~\citep{2004ApJ...616L.151R}. In order to quantitatively examine the influence of the propeller effect on the He-HMXB population, a better understanding of NS spin and magnetic field evolution is required, but this is not possible at present.
 

\subsection{CE and the Observed Be-HMXB population}
Since the He-HMXB population has both a sufficiently long lifespan and high luminosity to indicate the existence of numerous observable systems, another mechanism must halt the formation of these binaries. Since their progenitors are known to exist during the Be-HMXB phase, CEs stand as the only dynamical interaction which may eliminate He-HMXBs progenitors. Thus, it is possible to set constraints on the CE efficiency $\alpha_{CE}$ by demanding that enough of these systems are disrupted during a CE to bring the respective populations into agreement. Using the 13 Be-HMXBs with known orbital parameters as a template for the larger population of 81 systems and employing the detailed stellar evolution models calculated with STARS, we employ our orbital evolution code to determine the orbital and stellar parameters of the observed Be-HMXBs at the onset of the CE.

\begin{figure}
                \plotone{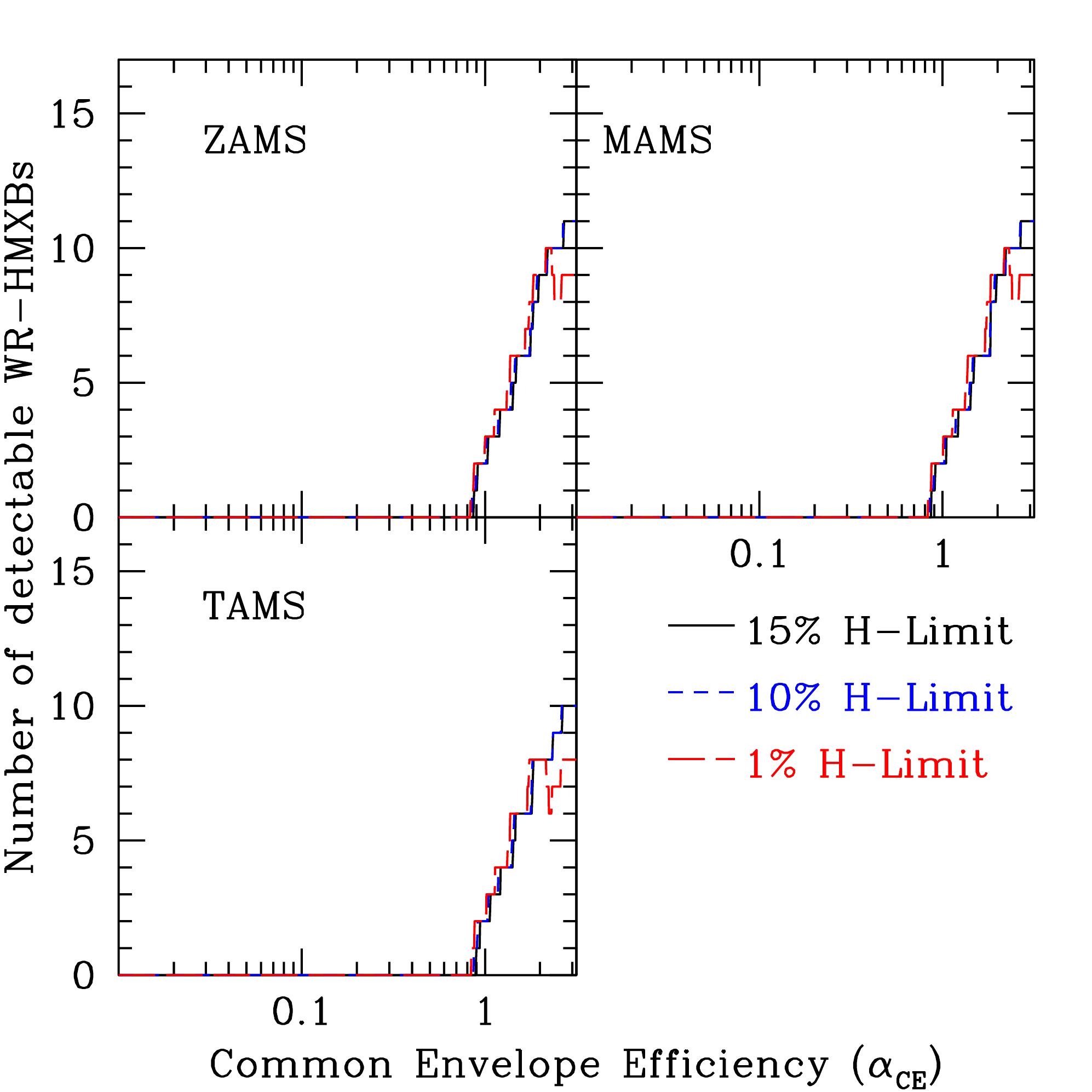}
                \caption{Number of surviving and observable (L$_x$~$>$~10$^{34}$~erg~s$^{-1}$) He-HMXBs as a function of the CE parameter $\alpha_{CE}$ employing the orbital characteristics at RLO determined via our stellar evolution code for the 13 known Be-HMXBs.}
\label{fig:cesurvivability}
\end{figure}

We again stress that we employ a calculation of $|E_{bind}|$ which includes not only the gravitational energy necessary to eject the primary envelope, but additionally the thermal energy in the envelope as well as H2 association and reionization. Since no other significant energy sources are available, the value of $\alpha_{CE}$ must fall below unity in order to conserve energy. In Figure~\ref{fig:cesurvivability}, we show the number of surviving He-HMXBs as a function of the CE efficiency $\alpha_{CE}$ under the assumption that detectable Be-HMXBs were observed at the ZAMS, MAMS and TAMS, and with  $X_{\rm{min}} $ set to 15\%, 10\%, and 1\%. We find nearly identical results in all scenarios, indicating that for $\alpha_{CE}$~$\lesssim$~1, mergers occur in all 13 of the observed Be-HMXB systems, regardless of their previous evolutionary history. Since one He-HMXB is in fact observed, and the lifetime of He-HMXBs may be smaller than that of Be-HMXBs, our model is only able to constrain $\alpha_{CE}$ to fall below unity, in line with theoretical expectations. We find that these results do indicate that current models are not missing any significant energy sources available to remove the CE, as the survivability of the CE phase exceeds 50\% for $\alpha_{CE}$~$>$~1.5, and these systems would be observable even with the weakened wind prescriptions of \citet{2002MNRAS.331.1027D}.

\section{A Complete Parameter Space}
\label{sec:grid}

In the previous sections we have shown that CE-driven mergers are necessary to explain the discrepancy between Be-HMXB and He-HMXB observations and that the typical CE parametrization, which demands an efficiency less than unity when all possible energy sources for unbinding the envelope are accounted for, is consistent with the observed sample of only one He-HMXB. However, these observed Be-HMXBs exist only as a subset for the potential class of He-HMXB progenitors. In addition to these luminous systems, there may exist a much larger population of binaries containing an O/B star and a NS in wide orbits, such that they are not bright X-ray systems. This underlying population remains undetected because either the donor star does not carry enhanced winds stemming from the Be phenomenon, or the system is too widely separated for stellar material to be effectively accreted onto the NS. In any case, the binary dynamics of this underlying population are identical to the visible population, and evolution through a CE phase may similarly result in bright He-HMXBs.

\begin{figure}
                \plotone{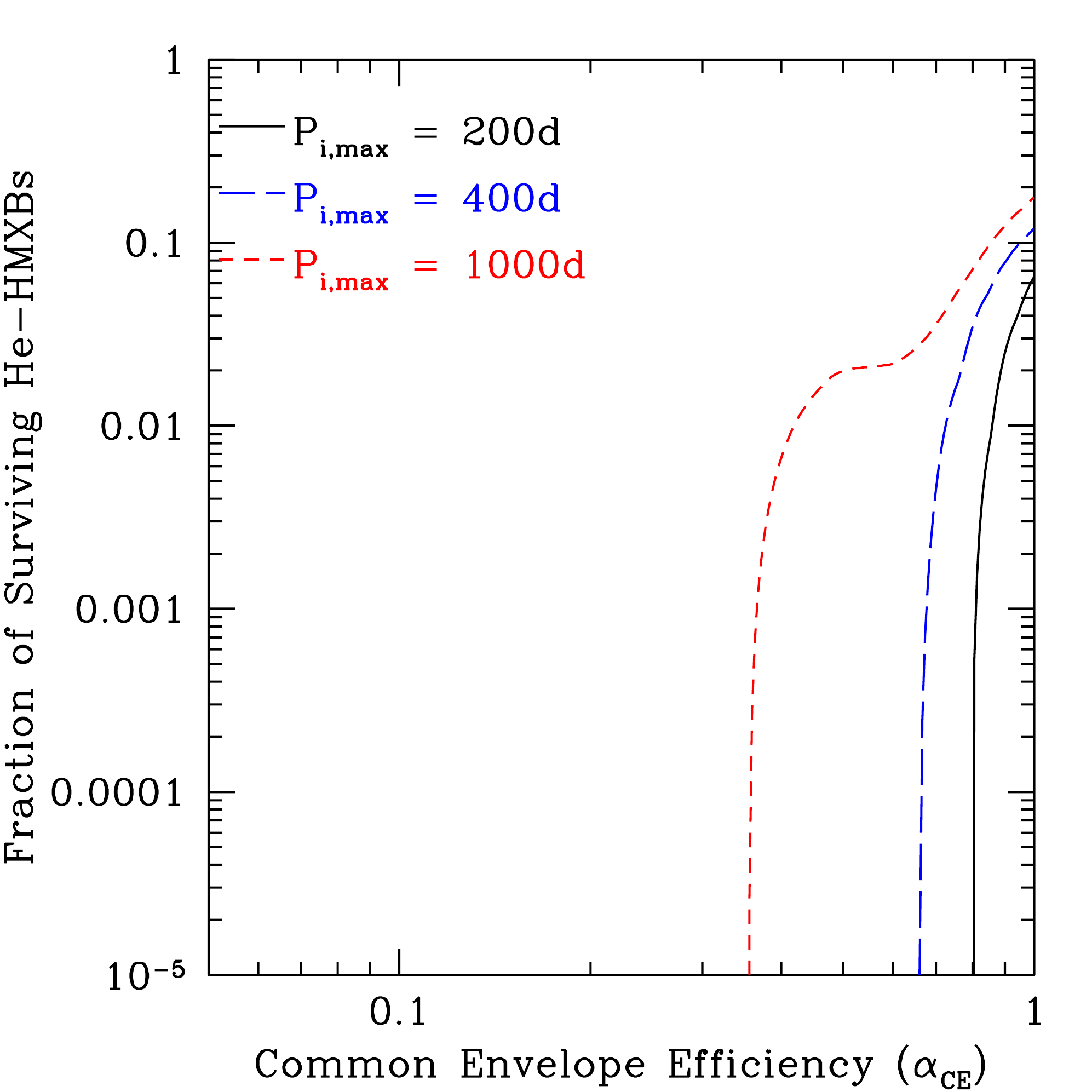}
                \caption{Fraction of surviving and visibly detected He-HMXBs from a parameter space of initially widely separated O/B-NS binaries defined in the text with a maximum initial orbital separation between the CO and MS star of 200 days (solid black), 400 days (blue long-dash) and 1000 days (red short-dash).}
\label{fig:grid}
\end{figure}

In order to model the evolution of these systems, we create a grid of binaries containing a NS and O/B donor following the parameters described in Section~\ref{sec:modeling}. In Figure~\ref{fig:grid}, we show the fraction of binaries which survives the CE yielding bright He-HMXBs. We assume maximum initial orbital periods of 200d (black solid), 400d (blue dashed), and 1000d (red dotted), following a distribution which is flat in the logarithm of the orbital period. We find that a potentially sizable fraction of He-HMXBs are created for larger values of $\alpha_{CE}$, although the bounds depend strongly on the maximum assumed orbital period of the underlying O/B-NS population. We note the observation of two Be-HMXBs with orbital periods above 200d allows us to set this as an observed lower limit for our simulated population (see Table~\ref{tab:behmxbs}). We also note that we do not expect a significant variation in our results if we were to vary the minimum orbital period in our sample grid (currently set at 10d). These relatively tightly bound systems are unlikely to survive a CE, and their inclusion in our models will not greatly affect the calculated number of He-HMXBs.

In order to apply these results to the expected number of observed He-HMXBs, we must normalize the number of systems in our grid against the expected number of loosely bound O/B-NS binaries. We note that the number of He-HMXBs expected from our simulated population of O/B-NS systems can be expressed as:

\begin{equation}
N_{He-HMXBs} = f_{s}(\frac{N_{Be}}{N_{B}})^{-1}(\frac{\tau_{He}}{\tau_{Be}})N_{O}
\end{equation}

where f$_s$ is the survival probability of a given system from our simulation grid (shown in Figure~\ref{fig:grid}), N$_{Be}$/N$_{B}$ provides the fraction of B-type donors which have Be-HMXB properties, $\tau_{He}$/$\tau_{Be-HMXB}$ describes the relative lifetime of the He-HMXB and Be-HMXB phases (shown in Figure~\ref{fig:relativetimescale}), and N$_{O}$ is the observed number of Be-HMXB systems (N$_O$~=~81 throughout this paper). 

The fraction of B-type stars which show emission-line (Be) spectra have been observed to vary between 2\%-7\%~\citep{2005ApJS..161..118M}, although we note some sources have shown Be-fractions as high as 8.5\%~\citep{2008ApJ...672..590M}. While this ratio may be substantially higher in binary systems if the spin-up of the Be population is due to binary interactions, this line of reasoning is disputed by \citet{2010MNRAS.405.2439O}, who find a similar binary fraction for both B and Be stars. In this work, we assume a Be-fraction of 7\%, and a fractional lifetime for the He-HMXB population ($\tau_{He}$/$\tau_{Be}$) of 20\%, taking a central value from Figure~\ref{fig:relativetimescale} under the assumption that the primary progenitor was not more than three times as massive as the Be-star. We note that the X-Ray detectability of these systems is evaluated for each surviving He-HMXB produced by our grid using the luminosity prescriptions of \citet{2002MNRAS.331.1027D} and a luminosity cutoff of 1$\times$10$^{34}$~erg~s$^{-1}$. From these values, we would anticipate a population of 230f$_s$ visible He-HMXBs. 

Thus, we constrain f$_s$ by comparing this expected population of He-HMXBs to the observation of only a single system. Noting that a prediction exceeding four He-HMXBs would create a 2$\sigma$  discrepancy with observation, we thus constrain the survivability of the CE phase to less than 2\%. Comparison with Figure~\ref{fig:grid} thus constrains the CE efficiency to $\alpha_{CE}$~$<$~0.88, $\alpha_{CE}$~$<$~0.75, and $\alpha_{CE}$~$<$~0.50 if the maximum orbital period is 200d, 400d, and 1000d, respectively.

We note that the above calculation is conservative in several ways. First, we have assumed that all systems containing a Be star and NS are visibly bright X-Ray sources. Secondly, we have assumed that the observed population of Be-HMXBs can be translated to a population of O/B-NS with an orbital period which is logarithmically flat starting at 10d. We note that the observed Be-HMXB population is instead biased towards systems with orbital periods around 100d. While the lack of observed loosely bound systems may be due to luminosity cutoffs or simply to limited to observational time, the low period population is likely complete. This implies that the survival fraction of O/B-NS systems resembling the Be-HMXBs may be substantially higher. However, less conservative estimations are unlikely to significantly alter the constraint imposed on the CE evolution, as Figure~\ref{fig:grid} shows that the survivability of the CE phase plunges for smaller values of $\alpha_{CE}$, implying that uncertainties in the estimation of the population of O/B-NS binaries has only a negligible effect on the number of expected He-HMXBs.

\section{Discussion and Conclusions}
\label{sec:conclusions}

Theoretical models predict the production of He-HMXBs through the CE evolution of widely separated binaries containing a NS and a massive donor. However, observations show a large population of Be-HMXBs and a significant lack of He-HMXBs. We find that detailed theoretical models predict the He-HMXB population to be sufficiently long-lived and luminous to be detected as the evolved offspring of the observed Be-HMXB population. Noting that a CE phase acts as the only dynamical mechanism which may disrupt the production of He-HMXBs, we use these observations to set constraints on the CE efficiency parameter $\alpha_{CE}$. Using the binary parameters of the observed Be-HMXBs, we are only able to limit $\alpha_{CE}$ to fall below unity, echoing theoretical constraints due to conservation of energy. Next, we simulate a larger grid of O/B-NS binaries with characteristics similar to the observed Be-HMXB population. From this grid, we constrain $\alpha_{CE}$ to be $<$0.88 for a population of O/B-NS binaries with a maximum orbital period of 200d, and possibly as low as $\alpha_{CE}$~$<$0.50 if the period extends to 1000d. We note that this extension of the O/B-NS population to high orbital periods may have observational support, as one X-Ray quiet system (B1259-63) with an orbital period of 1236d has been observed as a radio pulsar with an optically identified B-type companion~\citep{1999ApJ...522..504H, 1998MNRAS.298..997W}.

We note that the stringency of our constraints are limited primarily by the low number of observed Be-HMXBs, and especially by the limited number of Be-HMXBs with known orbital period and eccentricity information. We expect that observational detections of both new B-star NS binaries and determinations of the binary parameters of known Be-HMXBs will greatly reduce these measurement errors and provide a more accurate understanding of CE evolution in massive binaries. Furthermore, we note that our findings are complementary to previous studies which employ a combination of natal kicks and/or CE mergers to explain the small population of He-HMXBs with more massive (M$_{He}$~$>$5~M$_\odot$) Helium donors~\citep{1997A&A...318..812D, 1998NewA...3..443V, 2005A&A...443..231L}. Specifically, we extend the analysis to the lower mass range of He-HMXBs formed through the evolution of the Be-HMXB population and then use the number and binary properties of the observed Be-HMXBs to differentiate between the natal kick and CE merger hypotheses. We find that independent of any disruptions due to natal kicks (which would occur prior to the Be-HMXB phase), CE mergers must be common in order to explain the relative paucity of He-HMXBs. Using this, we can place strict limits on the CE efficiency.

Lastly, we note that a similar methodology may be applied to the population of known NS-NS binaries containing a pulsar, in order to determine whether the rate of NS-NS production is itself consistent with the low survivability probability assigned to Be-HMXBs moving through CE phases. Since these systems would additionally experience a He-HMXB phase in between the Be-HMXB and NS-NS phases, the existence of a large NS-NS population inconsistent with the small number of Be-HMXBs would instead point towards the existence of an X-Ray quiet population of He-HMXBs, and may be used as a further test of the results obtained here.

\acknowledgements We thank Ed Van den Heuvel for helpful comments during the development of this work. This work was partially supported by NSF grant AST--0908930 and NASA grant NNX10AH47G to VK. TL was supported in part by the Fermilab Fellowship in Theoretical Physics. Simulations were performed on the computing cluster {\tt Fugu} available to the Theoretical Astrophysics group at Northwestern and partially funded by NSF grant PHY--0619274 to VK. Finally we thank the organizers of the ESO Compact Object Binaries Conference where much of this research was completed. \\  \\

\bibliography{he}

\begin{thebibliography}{69}
\expandafter\ifx\csname natexlab\endcsname\relax\def\natexlab#1{#1}\fi

\bibitem[{{Abt}(1983)}]{1983ARA&A..21..343A}
{Abt}, H.~A. 1983, \araa, 21, 343

\bibitem[{{Bailyn} \& {Grindlay}(1987)}]{1987ApJ...316L..25B}
{Bailyn}, C.~D., \& {Grindlay}, J.~E. 1987, \apjl, 316, L25

\bibitem[{{Belczynski} {et~al.}(2008{\natexlab{a}}){Belczynski}, {Kalogera},
  {Rasio}, {Taam}, {Zezas}, {Bulik}, {Maccarone}, \&
  {Ivanova}}]{2008ApJS..174..223B}
{Belczynski}, K., {Kalogera}, V., {Rasio}, F.~A., {Taam}, R.~E., {Zezas}, A.,
  {Bulik}, T., {Maccarone}, T.~J., \& {Ivanova}, N. 2008{\natexlab{a}}, \apjs,
  174, 223, arXiv:astro-ph/0511811

\bibitem[{{Belczynski} {et~al.}(2008{\natexlab{b}}){Belczynski}, {Kalogera},
  {Rasio}, {Taam}, {Zezas}, {Bulik}, {Maccarone}, \&
  {Ivanova}}]{Belczynski2008}
------. 2008{\natexlab{b}}, Astrophys. J. Suppl. Ser., 174, 223

\bibitem[{{Belczynski} \& {Ziolkowski}(2009)}]{2009ApJ...707..870B}
{Belczynski}, K., \& {Ziolkowski}, J. 2009, \apj, 707, 870, 0907.4990

\bibitem[{{Bondi} \& {Hoyle}(1944)}]{BondiHoyle1944}
{Bondi}, H., \& {Hoyle}, F. 1944, Mon. Not. R. Astron. Soc., 104, 273

\bibitem[{{Chevalier} \& {Ilovaisky}(1998)}]{1998A&A...330..201C}
{Chevalier}, C., \& {Ilovaisky}, S.~A. 1998, \aap, 330, 201,
  arXiv:astro-ph/9710008

\bibitem[{{Claret}(2004)}]{2004A&A...424..919C}
{Claret}, A. 2004, \aap, 424, 919

\bibitem[{{Corbet} \& {Krimm}(2009)}]{2009ATel.2008....1C}
{Corbet}, R.~H.~D., \& {Krimm}, H.~A. 2009, The Astronomer's Telegram, 2008, 1

\bibitem[{{De Donder} {et~al.}(1997){De Donder}, {Vanbeveren}, \& {van
  Bever}}]{1997A&A...318..812D}
{De Donder}, E., {Vanbeveren}, D., \& {van Bever}, J. 1997, \aap, 318, 812

\bibitem[{{de Jager} {et~al.}(1988){de Jager}, {Nieuwenhuijzen}, \& {van der
  Hucht}}]{1988A&AS...72..259D}
{de Jager}, C., {Nieuwenhuijzen}, H., \& {van der Hucht}, K.~A. 1988, \aaps,
  72, 259

\bibitem[{{Dewi} {et~al.}(2002){Dewi}, {Pols}, {Savonije}, \& {van den
  Heuvel}}]{2002MNRAS.331.1027D}
{Dewi}, J.~D.~M., {Pols}, O.~R., {Savonije}, G.~J., \& {van den Heuvel},
  E.~P.~J. 2002, \mnras, 331, 1027, arXiv:astro-ph/0201239

\bibitem[{{Dewi} \& {Tauris}(2000)}]{2000A&A...360.1043D}
{Dewi}, J.~D.~M., \& {Tauris}, T.~M. 2000, \aap, 360, 1043,
  arXiv:astro-ph/0007034

\bibitem[{{Eggleton}(1971)}]{Eggleton1971}
{Eggleton}, P.~P. 1971, Mon. Not. R. Astron. Soc., 151, 351

\bibitem[{{Eggleton}(1973)}]{Eggleton1973}
------. 1973, Mon. Not. R. Astron. Soc., 163, 279

\bibitem[{{Eggleton} \& {Kiseleva-Eggleton}(2002)}]{Eggleton2002}
{Eggleton}, P.~P., \& {Kiseleva-Eggleton}, L. 2002, Astrophys. J., 575, 461

\bibitem[{{Eggleton} \& {Verbunt}(1986)}]{1986MNRAS.220P..13E}
{Eggleton}, P.~P., \& {Verbunt}, F. 1986, \mnras, 220, 13P

\bibitem[{{Habets} \& {Heintze}(1981)}]{1981A&AS...46..193H}
{Habets}, G.~M.~H.~J., \& {Heintze}, J.~R.~W. 1981, \aaps, 46, 193

\bibitem[{{Hamann} \& {Koesterke}(1998)}]{1998A&A...335.1003H}
{Hamann}, W.-R., \& {Koesterke}, L. 1998, \aap, 335, 1003

\bibitem[{{Hamann} {et~al.}(1995){Hamann}, {Koesterke}, \&
  {Wessolowski}}]{1995A&A...299..151H}
{Hamann}, W.-R., {Koesterke}, L., \& {Wessolowski}, U. 1995, \aap, 299, 151

\bibitem[{{Han} {et~al.}(1994){Han}, {Podsiadlowski}, \&
  {Eggleton}}]{1994MNRAS.270..121H}
{Han}, Z., {Podsiadlowski}, P., \& {Eggleton}, P.~P. 1994, \mnras, 270, 121

\bibitem[{{Han} {et~al.}(1995){Han}, {Podsiadlowski}, \&
  {Eggleton}}]{1995MNRAS.272..800H}
------. 1995, \mnras, 272, 800

\bibitem[{{Heger} {et~al.}(2003){Heger}, {Fryer}, {Woosley}, {Langer}, \&
  {Hartmann}}]{2003ApJ...591..288H}
{Heger}, A., {Fryer}, C.~L., {Woosley}, S.~E., {Langer}, N., \& {Hartmann},
  D.~H. 2003, \apj, 591, 288, arXiv:astro-ph/0212469

\bibitem[{{Heggie}(1975)}]{1975MNRAS.173..729H}
{Heggie}, D.~C. 1975, \mnras, 173, 729

\bibitem[{{Hjellming}(1989)}]{1989PhDT.........7H}
{Hjellming}, M.~S. 1989, PhD thesis, Illinois Univ.~at Urbana-Champaign, Savoy.

\bibitem[{{Hughes} \& {Bailes}(1999)}]{1999ApJ...522..504H}
{Hughes}, A., \& {Bailes}, M. 1999, \apj, 522, 504

\bibitem[{{Hurley} {et~al.}(2000){Hurley}, {Pols}, \&
  {Tout}}]{2000MNRAS.315..543H}
{Hurley}, J.~R., {Pols}, O.~R., \& {Tout}, C.~A. 2000, \mnras, 315, 543,
  arXiv:astro-ph/0001295

\bibitem[{{Hurley} {et~al.}(2002){Hurley}, {Tout}, \&
  {Pols}}]{2002MNRAS.329..897H}
{Hurley}, J.~R., {Tout}, C.~A., \& {Pols}, O.~R. 2002, \mnras, 329, 897,
  arXiv:astro-ph/0201220

\bibitem[{{Hut}(1981)}]{Hut1981}
{Hut}, P. 1981, Astron. Astrophys., 99, 126

\bibitem[{{Iben} \& {Livio}(1993)}]{1993PASP..105.1373I}
{Iben}, Jr., I., \& {Livio}, M. 1993, \pasp, 105, 1373

\bibitem[{{Ivanova} \& {Taam}(2004)}]{2004ApJ...601.1058I}
{Ivanova}, N., \& {Taam}, R.~E. 2004, \apj, 601, 1058, arXiv:astro-ph/0310126

\bibitem[{{Junker} \& {Schaefer}(1992)}]{Junker1992}
{Junker}, W., \& {Schaefer}, G. 1992, R. Astron. Soc., Mon. Not., 254, 146

\bibitem[{{Kaur} {et~al.}(2008){Kaur}, {Paul}, {Kumar}, \&
  {Sagar}}]{2008MNRAS.386.2253K}
{Kaur}, R., {Paul}, B., {Kumar}, B., \& {Sagar}, R. 2008, \mnras, 386, 2253,
  0803.1113

\bibitem[{{Linden} {et~al.}(2009){Linden}, {Sepinsky}, {Kalogera}, \&
  {Belczynski}}]{2009ApJ...699.1573L}
{Linden}, T., {Sepinsky}, J.~F., {Kalogera}, V., \& {Belczynski}, K. 2009,
  \apj, 699, 1573, 0807.1097

\bibitem[{{Lommen} {et~al.}(2005){Lommen}, {Yungelson}, {van den Heuvel},
  {Nelemans}, \& {Portegies Zwart}}]{2005A&A...443..231L}
{Lommen}, D., {Yungelson}, L., {van den Heuvel}, E., {Nelemans}, G., \&
  {Portegies Zwart}, S. 2005, \aap, 443, 231, arXiv:astro-ph/0507304

\bibitem[{{Loveridge} {et~al.}(2010){Loveridge}, {van der Sluys}, \&
  {Kalogera}}]{2010arXiv1009.5400L}
{Loveridge}, A.~J., {van der Sluys}, M., \& {Kalogera}, V. 2010, ArXiv
  e-prints, 1009.5400

\bibitem[{{McSwain} \& {Gies}(2005)}]{2005ApJS..161..118M}
{McSwain}, M.~V., \& {Gies}, D.~R. 2005, \apjs, 161, 118,
  arXiv:astro-ph/0505032

\bibitem[{{McSwain} {et~al.}(2008){McSwain}, {Huang}, {Gies}, {Grundstrom}, \&
  {Townsend}}]{2008ApJ...672..590M}
{McSwain}, M.~V., {Huang}, W., {Gies}, D.~R., {Grundstrom}, E.~D., \&
  {Townsend}, R.~H.~D. 2008, \apj, 672, 590, 0710.0137

\bibitem[{{Moffat} {et~al.}(1982{\natexlab{a}}){Moffat}, {Firmani}, {McLean},
  \& {Seggewiss}}]{1982IAUS...99..577M}
{Moffat}, A.~F.~J., {Firmani}, C., {McLean}, I.~S., \& {Seggewiss}, W.
  1982{\natexlab{a}}, in IAU Symposium, Vol.~99, Wolf-Rayet Stars:
  Observations, Physics, Evolution, ed. {C.~W.~H.~De Loore \& A.~J.~Willis},
  577--581

\bibitem[{{Moffat} {et~al.}(1982{\natexlab{b}}){Moffat}, {Lamontagne}, \&
  {Seggewiss}}]{1982A&A...114..135M}
{Moffat}, A.~F.~J., {Lamontagne}, R., \& {Seggewiss}, W. 1982{\natexlab{b}},
  \aap, 114, 135

\bibitem[{{Muno} {et~al.}(2006){Muno}, {Clark}, {Crowther}, {Dougherty}, {de
  Grijs}, {Law}, {McMillan}, {Morris}, {Negueruela}, {Pooley}, {Portegies
  Zwart}, \& {Yusef-Zadeh}}]{2006ApJ...636L..41M}
{Muno}, M.~P. {et~al.} 2006, \apjl, 636, L41, arXiv:astro-ph/0509408

\bibitem[{{Nebot G{\'o}mez-Mor{\'a}n} {et~al.}(2011){Nebot
  G{\'o}mez-Mor{\'a}n}, {G{\"a}nsicke}, {Schreiber}, {Rebassa-Mansergas},
  {Schwope}, {Southworth}, {Aungwerojwit}, {Bothe}, {Davis}, {Kolb},
  {M{\"u}ller}, {Papadaki}, {Pyrzas}, {Rabitz}, {Rodr{\'{\i}}guez-Gil},
  {Schmidtobreick}, {Schwarz}, {Tappert}, {Toloza}, {Vogel}, \&
  {Zorotovic}}]{2011arXiv1109.6662N}
{Nebot G{\'o}mez-Mor{\'a}n}, A. {et~al.} 2011, ArXiv e-prints, 1109.6662

\bibitem[{{Nelemans} \& {van den Heuvel}(2001)}]{2001A&A...376..950N}
{Nelemans}, G., \& {van den Heuvel}, E.~P.~J. 2001, \aap, 376, 950,
  arXiv:astro-ph/0107410

\bibitem[{{Nomoto}(1984)}]{1984ApJ...277..791N}
{Nomoto}, K. 1984, \apj, 277, 791

\bibitem[{{Nugis} \& {Lamers}(2000)}]{2000A&A...360..227N}
{Nugis}, T., \& {Lamers}, H.~J.~G.~L.~M. 2000, \aap, 360, 227

\bibitem[{{Oudmaijer} \& {Parr}(2010)}]{2010MNRAS.405.2439O}
{Oudmaijer}, R.~D., \& {Parr}, A.~M. 2010, \mnras, 405, 2439, 1003.0618

\bibitem[{{Paczynski}(1976)}]{1976IAUS...73...75P}
{Paczynski}, B. 1976, in IAU Symposium, Vol.~73, Structure and Evolution of
  Close Binary Systems, ed. {P.~Eggleton, S.~Mitton, \& J.~Whelan}, 75--+

\bibitem[{{Pols} {et~al.}(1995){Pols}, {Tout}, {Eggleton}, \&
  {Han}}]{Polsetal1995}
{Pols}, O.~R., {Tout}, C.~A., {Eggleton}, P.~P., \& {Han}, Z. 1995, Mon. Not.
  R. Astron. Soc., 274, 964

\bibitem[{{Raguzova} \& {Popov}(2005)}]{2005A&AT...24..151R}
{Raguzova}, N.~V., \& {Popov}, S.~B. 2005, Astronomical and Astrophysical
  Transactions, 24, 151, arXiv:astro-ph/0505275

\bibitem[{{Rasio} \& {Livio}(1996)}]{1996ApJ...471..366R}
{Rasio}, F.~A., \& {Livio}, M. 1996, \apj, 471, 366, arXiv:astro-ph/9511054

\bibitem[{{Ricker} \& {Taam}(2008)}]{2008ApJ...672L..41R}
{Ricker}, P.~M., \& {Taam}, R.~E. 2008, \apjl, 672, L41, 0710.3631

\bibitem[{{Rodriguez} {et~al.}(2010){Rodriguez}, {Tomsick}, \&
  {Bodaghee}}]{2010A&A...517A..14R}
{Rodriguez}, J., {Tomsick}, J.~A., \& {Bodaghee}, A. 2010, \aap, 517, A14+,
  1003.3741

\bibitem[{{Rodriguez} {et~al.}(2009){Rodriguez}, {Tomsick}, \&
  {Chaty}}]{2009A&A...494..417R}
{Rodriguez}, J., {Tomsick}, J.~A., \& {Chaty}, S. 2009, \aap, 494, 417,
  0811.4707

\bibitem[{{Romanova} {et~al.}(2004){Romanova}, {Ustyugova}, {Koldoba}, \&
  {Lovelace}}]{2004ApJ...616L.151R}
{Romanova}, M.~M., {Ustyugova}, G.~V., {Koldoba}, A.~V., \& {Lovelace},
  R.~V.~E. 2004, \apjl, 616, L151, arXiv:astro-ph/0502266

\bibitem[{{Salpeter}(1955)}]{1955ApJ...121..161S}
{Salpeter}, E.~E. 1955, \apj, 121, 161

\bibitem[{{Sandquist} {et~al.}(2000){Sandquist}, {Taam}, \&
  {Burkert}}]{2000ApJ...533..984S}
{Sandquist}, E.~L., {Taam}, R.~E., \& {Burkert}, A. 2000, \apj, 533, 984,
  arXiv:astro-ph/9912243

\bibitem[{{Sepinsky} {et~al.}(2007){Sepinsky}, {Willems}, \&
  {Kalogera}}]{SepinskyEtAlNoFred2007}
{Sepinsky}, J.~F., {Willems}, B., \& {Kalogera}, V. 2007, Astrophys. J., 660,
  1624

\bibitem[{{Taam} \& {Ricker}(2006)}]{2006astro.ph.11043T}
{Taam}, R.~E., \& {Ricker}, P.~M. 2006, ArXiv Astrophysics e-prints,
  arXiv:astro-ph/0611043

\bibitem[{{Taam} \& {Sandquist}(2000)}]{2000ARA&A..38..113T}
{Taam}, R.~E., \& {Sandquist}, E.~L. 2000, \araa, 38, 113

\bibitem[{{Van Bever} \& {Vanbeveren}(2000)}]{2000A&A...358..462V}
{Van Bever}, J., \& {Vanbeveren}, D. 2000, \aap, 358, 462

\bibitem[{{van den Heuvel}(1976)}]{1976IAUS...73...35V}
{van den Heuvel}, E.~P.~J. 1976, in IAU Symposium, Vol.~73, Structure and
  Evolution of Close Binary Systems, ed. {P.~Eggleton, S.~Mitton, \&
  J.~Whelan}, 35--+

\bibitem[{{van den Heuvel} {et~al.}(2000){van den Heuvel}, {Portegies Zwart},
  {Bhattacharya}, \& {Kaper}}]{2000A&A...364..563V}
{van den Heuvel}, E.~P.~J., {Portegies Zwart}, S.~F., {Bhattacharya}, D., \&
  {Kaper}, L. 2000, \aap, 364, 563, arXiv:astro-ph/0005245

\bibitem[{{van Kerkwijk} {et~al.}(1992){van Kerkwijk}, {Charles}, {Geballe},
  {King}, {Miley}, {Molnar}, {van den Heuvel}, {van der Klis}, \& {van
  Paradijs}}]{1992Natur.355..703V}
{van Kerkwijk}, M.~H. {et~al.} 1992, \nat, 355, 703

\bibitem[{{Vanbeveren} {et~al.}(1998{\natexlab{a}}){Vanbeveren}, {De Donder},
  {van Bever}, {van Rensbergen}, \& {De Loore}}]{1998NewA...3..443V}
{Vanbeveren}, D., {De Donder}, E., {van Bever}, J., {van Rensbergen}, W., \&
  {De Loore}, C. 1998{\natexlab{a}}, 3, 443

\bibitem[{{Vanbeveren} {et~al.}(1998{\natexlab{b}}){Vanbeveren}, {De Loore}, \&
  {Van Rensbergen}}]{1998A&ARv...9...63V}
{Vanbeveren}, D., {De Loore}, C., \& {Van Rensbergen}, W. 1998{\natexlab{b}},
  \aapr, 9, 63

\bibitem[{{Webbink}(1984)}]{1984ApJ...277..355W}
{Webbink}, R.~F. 1984, \apj, 277, 355

\bibitem[{{Wex} {et~al.}(1998){Wex}, {Johnston}, {Manchester}, {Lyne},
  {Stappers}, \& {Bailes}}]{1998MNRAS.298..997W}
{Wex}, N., {Johnston}, S., {Manchester}, R.~N., {Lyne}, A.~G., {Stappers},
  B.~W., \& {Bailes}, M. 1998, \mnras, 298, 997, arXiv:astro-ph/9803182

\bibitem[{{Zahn}(1977)}]{Zahn1977}
{Zahn}, J. 1977, Astron. Astrophys., 57, 383

\bibitem[{{Zahn}(1989)}]{Zahn1989}
------. 1989, Astron. Astrophys., 220, 112

\end{thebibliography}
\end{document}